\documentclass[10pt, aps, prl, amssymb, amsmath, amsfonts, showpacs, floatfix, twocolumn, twoside, a4paper, superscriptaddress, longbibliography, nofootinbib]{revtex4-2}

\usepackage{graphicx}
\usepackage[usenames,dvipsnames]{xcolor} 
\usepackage[utf8]{inputenc}
\usepackage{color}
\usepackage{hyperref}
\usepackage[normalem]{ulem} 
\usepackage{physics}
\usepackage[capitalise]{cleveref}
\usepackage{mathrsfs, mathtools}
\usepackage{siunitx}
\usepackage{comment}
\usepackage{soul}
\usepackage[T3,T1]{fontenc}
\DeclareMathAlphabet{\mathpzc}{OT1}{pzc}{m}{it} 
\usepackage{mathrsfs} 
\usepackage{bm} 
\usepackage{microtype} 
\def\MT@register@subst@font{
	\MT@exp@one@n\MT@in@clist\font@name\MT@font@list
	\ifMT@inlist@\else\xdef\MT@font@list{\MT@font@list\font@name,}\fi}
\makeatother
\usepackage{orcidlink}
\usepackage{multirow}
\usepackage{booktabs} 
\usepackage{array} 
\usepackage{tabularx}
\usepackage{braket} 
\usepackage{dblfloatfix}

\newcolumntype{C}{>{\centering\arraybackslash}X}

\usepackage{hyperref} 
\usepackage{nicefrac} 

\usepackage{enumitem}
\usepackage{comment}
\usepackage{comment}
\usepackage{xspace}
\usepackage{bm}
\hypersetup{
  colorlinks   = true, 
  urlcolor     = blue, 
  linkcolor    = blue, 
  citecolor   = blue 
}
\allowdisplaybreaks[1]
\graphicspath{{figures/}}

\usepackage{float}

\renewcommand{\dd}{\mathrm{d}}
\newcommand{\mb}{m_\phi}

\newcommand{\yrs}{\text{yrs}}
\newcommand{\Hz}{\ensuremath{\,\mathrm{Hz}}\xspace}

\newcommand{\phXP}{\texttt{IMRPhenomXP}\xspace}

\newcommand{\phXPHM}{\texttt{IMRPhenomXPHM}\xspace}

\newcommand{\bilby}{\textsc{Bilby}\xspace}

\newcommand{\lalsuite}{\textsc{LALSuite}\xspace}

\newcommand{\dynesty}{\textsc{Dynesty}\xspace}
\newcommand{\numpy}{\textsc{NumPy}\xspace}
\newcommand{\scipy}{\textsc{SciPy}\xspace}
\newcommand{\matplotlib}{\textsc{Matplotlib}\xspace}
\newcommand{\arviz}{\textsc{ArviZ}\xspace}

\newcommand{\niceunderscore}{\raisebox{0.3ex}{\scalebox{0.7}[0.8]{\_}}}

\def \densityUnit {\,\si{\gram\per\cubic\centi\meter}}
\def \lnB {\ln\mathcal{B}_{\rm vac}^{\rm env}}
\def \rhobar {\ensuremath{\bar{\rho}_\phi}\xspace}

\newcommand{\mitctp}{Center for Theoretical Physics, Massachusetts Institute of Technology, Cambridge, MA 02139, USA}
\newcommand{\louvain}{Center for Cosmology, Particle Physics and Phenomenology - CP3, Université Catholique de Louvain, Louvain-La-Neuve, B-1348, Belgium}
\newcommand{\belgium}{Royal Observatory of Belgium, Avenue Circulaire, 3, 1180 Uccle, Belgium}
\newcommand{\grappa}{Gravitation Astroparticle Physics Amsterdam (GRAPPA), University of Amsterdam, Science Park 904, 1098 XH, Amsterdam, The Netherlands}
\newcommand{\oxford}{Astrophysics, University of Oxford, DWB, Keble Road, Oxford OX1 3RH, United Kingdom}
\newcommand{\qmul}{Geometry, Analysis and Gravitation, School of Mathematical Sciences, Queen Mary University of London,
Mile End Road, London E1 4NS, United Kingdom}

\begin{document}

\title{Scalar fields around black hole binaries in LIGO-Virgo-KAGRA}

\author{Soumen Roy}
\email{soumen.roy@uclouvain.be}
\affiliation{\louvain}
\affiliation{\belgium}
\author{Rodrigo Vicente}
\email{r.l.lourencovicente@uva.nl}
\affiliation{\grappa}
\author{Josu C. Aurrekoetxea}
\email{jaurreko@mit.edu}
\affiliation{\mitctp}
\author{Katy Clough}
\affiliation{\qmul}
\author{Pedro G. Ferreira}
\affiliation{\oxford}

\begin{abstract}
Light scalar particles arise naturally in many extensions of the Standard Model and are compelling dark-matter candidates. Gravitational interactions near black holes can trigger the growth of dense scalar configurations that, if sustained during inspiral, alter binary dynamics and imprint signatures on gravitational-wave signals. Detecting such effects would provide a novel probe of fundamental physics and dark matter. Here we develop a semi-analytic waveform model for binaries in scalar environments, validate it against numerical relativity simulations, and apply it in a Bayesian analysis of the LIGO–Virgo–KAGRA catalog.
We obtain physically meaningful upper limits on scalar densities around most compact binaries. For GW190728 and GW190814, vacuum lies outside the $95\%$ credible region. When including superradiance priors, GW190728 shows tentative evidence for a scalar environment with a Bayes factor of $\ln \mathcal{B}^{\rm env}_{\rm vac} \approx 3.5$, consistent with a light scalar of mass~$\sim10^{-12}\,\mathrm{eV}$.
\end{abstract}

\maketitle

{\em \textbf{Introduction.}---}
Gravitational waves (GWs) from compact binary coalescences provide a unique probe of the environments of black holes (BHs)~\cite{Barausse:2014tra, Tamanini:2019usx, Cardoso:2019rou, CanevaSantoro:2023aol, Zwick:2025wkt, Toubiana:2020drf, Derdzinski:2020wlw, Zwick:2021dlg, Sberna:2022qbn, Vijaykumar:2023tjg, Roy:2024rhe, DuttaRoy:2025gnu, Duque:2025yfm, HegadeKR:2025dur, HegadeKR:2025rpr}, allowing us to test the density of dark matter (DM) on small scales and potentially constrain its microphysics~\cite{Barack:2018yly, Bertone:2019irm, Annulli:2020lyc, Kavanagh:2020cfn, Maselli:2021men, Coogan:2021uqv, Vicente:2022ivh, Traykova:2023qyv, Cardoso:2021wlq, Cardoso:2022whc, Spieksma:2024voy, Speeney:2024mas, Pezzella:2024tkf, Gliorio:2025cbh, Cole:2022yzw, Baumann:2021fkf, 2307.16093, Duque:2023seg, Tomaselli:2023ysb, Tomaselli:2024bdd, Tomaselli:2024dbw, Boskovic:2024fga, Dyson:2025dlj, Rahman:2023sof, Karydas:2024fcn, Kavanagh:2024lgq, Spieksma:2024voy, Blas:2024duy, Miller:2025yyx, Bertone:2024rxe, DellaMonica:2025zby, Leong:2023nuk, Vicente:2025gsg}. If DM interacts only gravitationally, GW observations may be the only channel through which we can test its nature.

Light bosons, well-motivated as DM candidates~\cite{Arvanitaki:2009fg, Hui:2016ltb, Hui:2021tkt, Ferreira:2020fam}, can form dense configurations around BHs through several mechanisms, such as superradiance~\cite{1972Natur.238..211P, 1971JETPL..14..180Z, Detweiler:1980uk, Zouros:1979iw, Cardoso:2004nk, Dolan:2007mj, East:2017ovw, East:2018glu} (see \cite{Brito:2015oca} for a review),
dynamical capture via self-interactions~\cite{Budker:2023sex}, binary-induced bound states~\cite{Wong:2020qom, Bamber:2022pbs, Aurrekoetxea:2023jwk,Aurrekoetxea:2024cqd,Tomaselli:2024ojz,Srikanth:2025lic}, and accretion-driven spikes~\cite{Hui:2019aqm, Clough:2019jpm, Bamber:2020bpu, Cardoso:2022nzc, Hancock:2025ois}.
Superradiant clouds can have up to $10\%$ of the BH mass~\cite{East:2018glu, Herdeiro:2021znw}, reaching average densities above $10^9\densityUnit$ around stellar-mass BHs---over 30 orders of magnitude above the galactic DM background. Such densities lie well within the reach of LIGO--Virgo--KAGRA (LVK)~\cite{LIGOScientific:2014pky, VIRGO:2014yos, KAGRA:2013rdx, Virgo:2022ysc, KAGRA:2020tym, aLIGO:2020wna}, which has already constrained putative gaseous environments at the level of $\sim10\densityUnit$~\cite{CanevaSantoro:2023aol, Roy:2024rhe}.

It has long been suggested that close-to-equal-mass binaries---the primary targets of LVK~\cite{2025arXiv250818083T}---would disrupt these DM structures, based on studies of \emph{spikes} of heavier particle DM candidates~\cite{Merritt:2003qk, Bertone:2005hw, Kavanagh:2018ggo, Cole:2022ucw}. As a result, extreme mass-ratio inspirals (to be observed soon by LISA~\cite{LISA:2022kgy, LISA:2024hlh} and TianQin~\cite{TianQin:2015yph}) have been considered more suitable targets to detect such environments around the primary object~\cite{Baumann:2021fkf, Tomaselli:2023ysb, 2307.16093, Duque:2023seg, Dyson:2025dlj}.
However, numerical relativity (NR) simulations of scalar fields around equal-mass BH binaries~\cite{Ikeda:2020xvt, Bamber:2022pbs, Aurrekoetxea:2023jwk, Aurrekoetxea:2024cqd, Xin:2025ymm,Machet:2025vzt,Ficarra:2021qeh,Cheng:2025wac} and several different analytic approaches~\cite{Wong:2020qom, Tomaselli:2024ojz} have suggested that not only can a considerable fraction of an existing DM structure survive, but additional mass may also be captured by binary dynamics, with the formation of bound states from infalling matter.
These findings motivate a dedicated search for the effects of light scalars in the GW events from LVK. 

In this Letter, we present a fast semi-analytic waveform model for BH binaries in scalar environments and validate it against NR simulations. Applying it in a Bayesian analysis of the GWTC-3 catalog~\cite{LIGOScientific:2018mvr, LIGOScientific:2020ibl, LIGOScientific:2021usb, KAGRA:2021vkt}, we obtain the first upper bounds on scalar-field environments around compact binaries. When including superradiance priors, GW190728{\niceunderscore}064510~\cite{LIGOScientific:2020ibl} (hereafter GW190728) shows tentative evidence for a scalar environment with a Bayes factor of $\lnB \approx 3.5$. If confirmed, this would point to a new light scalar with mass $\mb \sim 10^{-12}\,\mathrm{eV}$.

\vskip 5pt
{\em \textbf{Scalars around black holes.}---}
We consider a real scalar field~$\phi$ minimally coupled to gravity with mass $\mb$, yielding the Einstein-Klein-Gordon (EKG) system of equations (we use $G=c=\hbar=1$)
\begin{equation} \label{eq:EKG}
	G_{\mu \nu}[g]=8\pi  T^\phi_{\mu \nu}\,, \qquad \left[\Box_g -\mb^2\right]\phi=0\,,
\end{equation}
where $G_{\mu \nu}$ is the Einstein tensor and $T^\phi_{\;\mu \nu}$ is the energy-momentum tensor of the scalar field $\phi$.

Our model will assume that some scalar field remains around the binary in the late stages of the inspiral. 
Several works (e.g.,~\cite{Baumann:2021fkf, Tomaselli:2023ysb, 2307.16093, Duque:2023seg, Tomaselli:2024bdd, Tomaselli:2024dbw, Boskovic:2024fga, Dyson:2025dlj}) have applied a perturbative approach in the parameter $\epsilon \equiv r_\phi q / \max(R,r_\phi)\ll 1$, with $r_\phi\equiv M/\alpha^2$, $\alpha\equiv \mb M$, and $q\equiv m_2/m_1$, to study the gravitational interaction of a companion of mass $m_2$ at a separation $R$ with a scalar field structure of superradiant origin (a \textit{gravitational atom} state) around a BH of mass $m_1$, and its consequent effect on binary dynamics. In that regime, it was shown in~\cite{Tomaselli:2024bdd, Tomaselli:2024dbw} that superradiant clouds can survive up to the late inspiral only for a particular region of parameter space.
However, much less is known about close-to-equal-mass binaries, where backreaction~\cite{Takahashi:2021yhy} and astrophysical processes~\cite{Guo:2024iye} (e.g., a common envelope phase) may help the cloud survive induced resonant absorption, since the perturbative scheme in $\epsilon$ breaks down in the late inspiral of compact binaries with $q\sim 1$ and small orbital separations $R\lesssim r_\phi$.

Our model does not rely on a specific scalar configuration; it only requires significant support in the monopole $\ell = 0$ \textit{gravitational molecule} states during the late inspiral. 
To construct our semi-analytic model, we employ a perturbative scheme tailored to near-equal-mass binaries (similar to approaches in~\cite{Ikeda:2020xvt, Tomaselli:2024ojz, Guo:2025pea}), which we now summarize.

First, we take the non-relativistic limit. We assume that (i) the scalars evolve mostly in the weak gravitational field $(g_{\mu \nu})\approx \mathrm{diag}[-(1-2 U),(1+2U)\delta_{ij}]$ with $U\ll1$, and (ii) the scalar field has most of its support on non-relativistic modes, i.e., $|\partial_t\log\psi|/\mb\sim U$, using the field redefinition~$\phi\equiv \psi\, e^{-i \mb t}/\sqrt{2 \mb}  + {\rm c.c.}$. The EKG system for $(g_{\mu \nu},\phi)$, then reduces to the much simpler Poisson-Schrödinger system for $(U,\psi)$. In most relevant scenarios, the self-gravity of the scalars is negligible, and
\begin{equation*}
    U\approx \sum_{\ell_*,m_*} \frac{4\pi M}{2 \ell_* +1} Y^*_{\ell_*m_*}(\theta_{2*}, \varphi_{2*}) Y_{\ell_*m_*}(\theta,\varphi)\mathcal{F}_{\ell_*}(r,r_{2*})
\end{equation*}
is simply the binary potential, in center of mass coordinates, with $(r_{2*},\theta_{2_*},\varphi_{2_*}(t))$ the position of the secondary, and $M$ the total binary mass; the explicit expression of $\mathcal{F}_{\ell_*}(r,r_{2*})$ is given in Supplemental Material (SM)~\cite{SupplementalMaterial}. 

Note that most scalars interacting effectively with the binary are near $r\sim r_\phi \equiv M/\alpha^2$, as it is impossible to sustain a long-lived structure with a smaller radius~\cite{Hui:2016ltb}. At such radii, $M F_{\ell _*}\propto \alpha^2 \beta^\ell$, where $\beta\equiv R/r_\phi$. For binaries in LVK, $\beta\sim 0.04(\alpha/0.06)^2[(100\mathrm{Hz}/f)(20M_\odot/M)]^{\frac{2}{3}}$, with $f\equiv \Omega/\pi$ the GW frequency of the dominant mode, and $\Omega$ the orbital (angular) frequency. This implies a hierarchy in the contribution of the binary multipoles to the Schrödinger equation and motivates a perturbative approach where, at zero order, the field is expressed as a sum of global (bound and scattering) molecule states of the binary monopole, which is then perturbatively scattered by the higher multipoles of the binary~\cite{Ikeda:2020xvt, Tomaselli:2024ojz, Guo:2025pea, Guo:2025ckp}.

The above perturbative scheme allows us to express the angular momentum exchange of the binary with the surrounding scalar field as
\begin{equation}\label{eq:torq}
    \frac{\dot{L}_\phi}{M}\approx - 4\pi \rhobar M^2 \frac{\mathcal{I}(v,\alpha,q)}{v^4}\,,
\end{equation}
where $v\equiv (\Omega M)^{1/3}$, $\rhobar$ is the average mass density of scalars in the region $r\lesssim r_\phi$, and $\mathcal{I}$ a simple analytical function of $v$, $\alpha$, and $q$ (for details, see the SM~\cite{SupplementalMaterial}). As shown in the following sections, whilst the model makes a number of strong assumptions, including being Newtonian, it is nevertheless sufficient to capture the changes in the relativistic dynamics observed in full NR simulations. 

\begin{figure}[t]
\href{https://youtu.be/9yTvjDl9L9A}{
\includegraphics[width=\columnwidth]{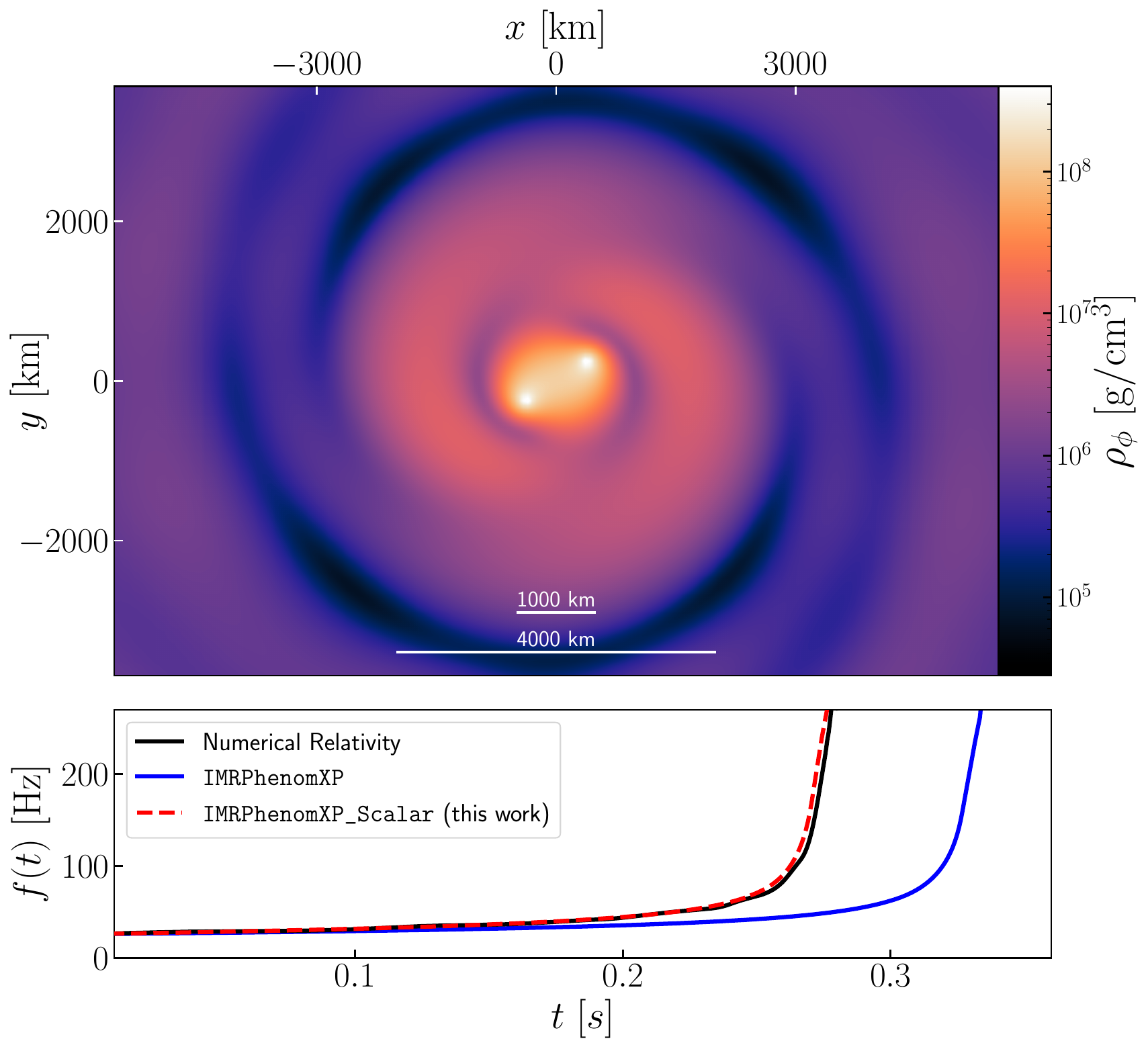}}
\caption{\textbf{Top:} Density snapshot of a NR simulation of an equal-mass binary BH with total mass $M=60 M_\odot$ in a scalar field with $\alpha=0.43$ and an asymptotic density $\rho_\phi \approx 8.6 \times 10^5\densityUnit$ at the boundaries of a cubic box with $L\approx4\times 10^4\,{\rm km}$. 
We can see a monopole overdense structure of radius {$r_\phi\equiv M/\alpha^2 \approx 480 \,{\rm km}$}; these develop around the binary in just a couple of orbits. \textbf{Bottom:} Waveform frequency extracted from an NR simulation compared to the maximum-likelihood samples from \phXP and our model \texttt{IMRPhenomXP\_Scalar} (see \cref{fig:pe:inj1}).}\label{fig:waveform_comparison}
\end{figure}

\vskip 5pt
{\em \textbf{Waveform model.}---}
The inspiral stage of close compact binaries probed by LVK is dominated by GW radiation reaction; for realistic mass densities \rhobar, the energy and angular momentum exchange with the surrounding scalar field is responsible for a perturbative dephasing relative to the vacuum chirp (bottom panel of Fig.~\ref{fig:waveform_comparison}). In particular, the late inspiral stage is expected to be quasi-circular, meaning that the torque on the binary (in \cref{eq:torq}) fully captures the effect of the environment on the dynamics. 

Our waveform model is built as in~\cite{Roy:2024rhe}, but here we include higher modes, which add considerable constraining power to the more asymmetric binary systems. We obtain $\dot{\Omega}(v)$ from the conservation of angular momentum, $\dot{L}=-(\dot{L}_{\rm GW}+\dot{L}_{\phi})$. Then, we compute the Fourier domain correction for the phase using the stationary phase approximation (SPA)~\cite{Cutler:1994ys, Droz:1999qx}. According to the SPA, the waveform modes are approximated as
\begin{equation}
\tilde{h}_{\ell m}(f) \approx A_{\ell m}(\tilde{v}) \sqrt{\frac{2\pi}{m \dot{\Omega}(\tilde{v})}} \: e^{i \left( 2\pi f t_c - \Phi_{\ell m} - \frac{\pi}{4} + \Phi_c \right)},
\end{equation}
where $t_c$ and $\Phi_c$ are the reference time and phase. The amplitude is given by $A_{\ell m} = 2 \sqrt{16 \pi/5}\, \tilde{v}^2\eta \,\hat{h}_{\ell m} $, with symmetric mass-ratio $\eta\equiv m_1 m_2/M^2$ and $\tilde{v} \equiv (2\pi M f / m)^{1/3}$; the expressions for $\hat{h}_{\ell m}$ up to 2PN order for non-spinning binaries can be found in \cite{Kidder:2007rt}. In the inspiral, the phase of different modes is related by $\Phi_{\ell m}(f) \approx (m/2) \Phi_{22}(2 f/ m)$, where $\Phi_{22}(f) = 2\pi f t(f) - \Phi(f)$ (e.g., \cite{Roy:2024rhe}). {Using this prescription, we incorporate the environmental corrections into the co-precessing frame of the \phXPHM model~\cite{Garcia-Quiros:2020qpx, Pratten:2020ceb}, as implemented in \lalsuite~\cite{lalsuite}; this is a phenomenological inspiral-merger-ringdown waveform model that includes precession and higher-order modes. For the comparison with the NR simulations, we used \phXP (dominant mode only) as the baseline vacuum model.
We label our environment waveform models with a suffix \texttt{\_Scalar}.}

\vskip 5pt
{\em \textbf{Validation of waveform model.}---}
%
We have generated waveforms directly from NR simulations to validate our waveform model. We used the \texttt{GRChombo} code~\cite{Andrade:2021rbd} to evolve the binary and scalar field system during a 10-orbit equal-mass quasi-circular BH coalescence, using an integrated version of \texttt{TwoPunctures}~\cite{Ansorg:2004ds} to obtain initial data for the vacuum binary system, and setting the initial trace of the extrinsic curvature tensor based on the asymptotic density of the scalar field to satisfy the constraints, as in~\cite{Aurrekoetxea:2022mpw, Aurrekoetxea:2025kmm, Aurrekoetxea:2023jwk, Aurrekoetxea:2024cqd}. We evolve the coupled EKG system using the CCZ4 formulation of Einstein's equations~\cite{Alic:2011gg} with the moving puncture gauge~\cite{Bona:1994dr, Baker:2005vv, Campanelli:2005dd, vanMeter:2006vi}. Finally, we extract and decompose the Newman-Penrose scalars into spherical harmonics, which we integrate using the fixed frequency integration method \cite{Reisswig:2010di} to obtain the strain waveforms~$h(t)$ in the time domain.

\Cref{fig:waveform_comparison} displays one of these NR simulations. The top panel shows a density snapshot for an equal-mass BBH with total mass $M=60 M_\odot$ in a scalar field environment with $\alpha=0.43$. In the bottom panel, the NR waveform is compared to the maximum-likelihood samples of the vacuum model \phXP and our scalar field model \texttt{IMRPhenomXP\_Scalar}, for a signal starting at a frequency of $26 \mathrm{\,Hz}$. Unlike the vacuum model, the scalar field one reproduces the frequency chirp seen in the NR simulations. 

\begin{figure}[t]
\includegraphics[width=\columnwidth]{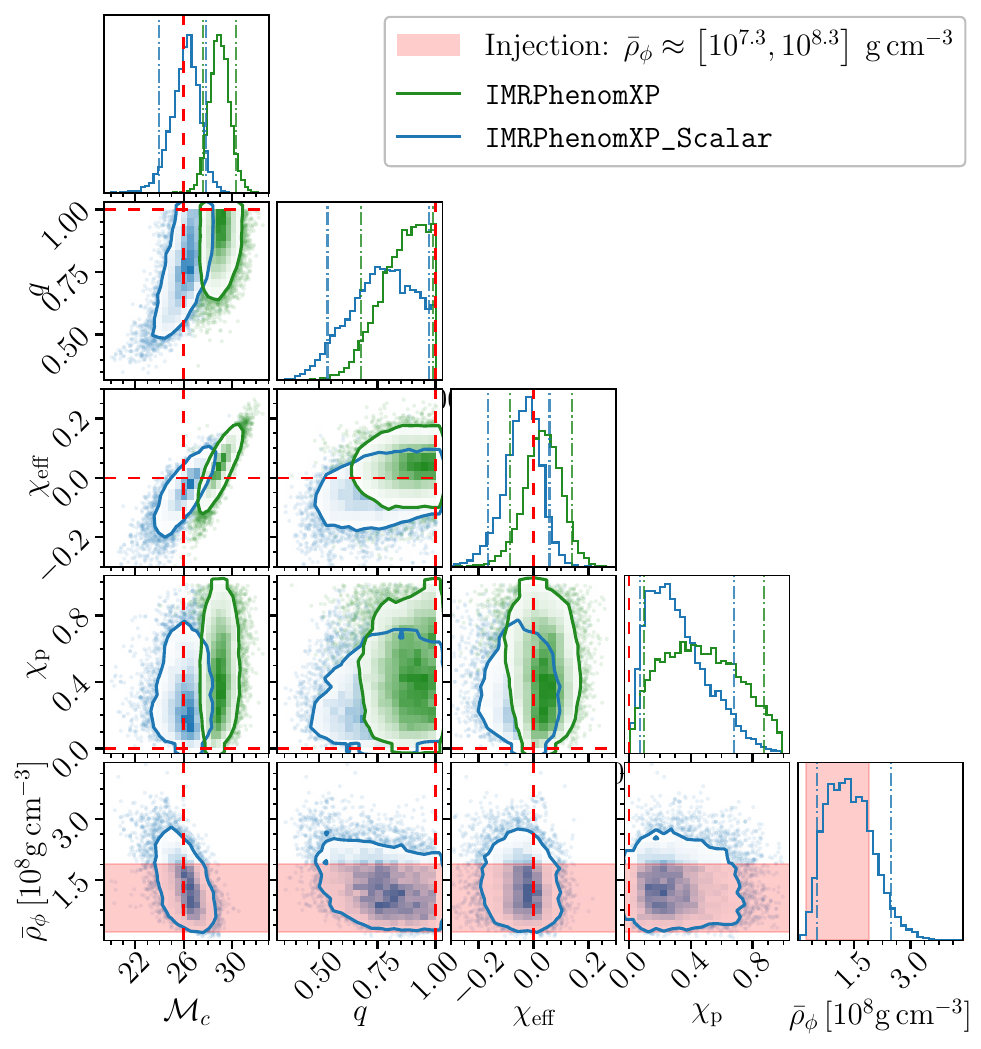}
\caption{Marginalized posterior distributions from the analyses of an injected NR waveform (same as in \cref{fig:waveform_comparison}), using the recovery vacuum model \phXP (green) and environment model \texttt{IMRPhenomXP\_Scalar} (blue). The red dashed lines indicate the injected values. The vertical dashed-dot lines in 1D histograms represent the 90\% credible interval. Unlike the vacuum model, which leads to significant biases, the environment model can accurately recover the system parameters; $\alpha$~is not well measurable and so its posterior is omitted. The Bayes factor in the model comparison is $\lnB=3.8$.}\label{fig:pe:inj1}
\end{figure}

To assess our model, we performed Bayesian parameter estimation on the injected NR waveforms using both models. The injections assumed zero noise, weighted by the GW150914 event PSDs of Hanford and Livingston detectors~\cite{LIGOScientific:2016aoc}, and an $\text{SNR} \approx 24$.
\Cref{fig:pe:inj1} shows the results for the injected NR waveform of \cref{fig:waveform_comparison}, using the recovery models \phXP and \texttt{IMRPhenomXP\_Scalar}.
The vacuum model fails to recover the true binary parameters: the estimation of the chirp mass is significantly biased to larger values, compensating for the accelerated inspiral caused by energy exchange with the environment. By contrast, our scalar field model resolves these biases and correctly infers the order of magnitude of the average scalar field density around the binary. The Bayes factor, $\lnB=3.8$, demonstrates the ability of our model to provide evidence of interaction with scalars. We validated the model for different asymptotic scalar field densities; another example is shown in Fig.~\ref{fig:pe:inj2} of the SM~\cite{SupplementalMaterial}.

In applying our Newtonian model to the NR waveform of \cref{fig:pe:inj1}, we extrapolate well beyond its regime of validity, and its use should therefore be treated with some caution. 
In addition to applying a Newtonian model in the relativistic regime, another issue is that the NR simulation does not provide a precise value for the parameter \rhobar that appears in \cref{eq:torq}, since the simulated profile is more complex than that of the model. We therefore simply check that its inferred value is comparable to the average over a sphere of radius $r_\phi$ during several orbits of the simulation. We find that the model provides the correct order of magnitude, but as a result the density should not be treated as a precise value.
Our model should become more reliable for systems with~$\alpha\ll1$, where the larger cloud size and smoother profile mitigate these issues. In this small $\alpha$ regime, the exchange of energy with the scalar field environment is more strongly suppressed, and is unlikely to leave a detectable imprint on a 10-orbit waveform \cite{Aurrekoetxea:2023jwk}. Therefore to validate it, we would need to follow the evolution of the system for a much higher number of orbits, which is computationally expensive using NR. We therefore need to rely on our validation in the high $\alpha$ limit, which seems reasonable since it should be the worst case scenario for accuracy of the model.
Injections with our phenomenological model \texttt{IMRPhenomXP\_Scalar} indicate that, for longer signals from lower-mass binaries (such as some LVK events analyzed below), the posterior on $\alpha$ becomes informative (consistent with~\cite{DellaMonica:2025zby}) and the Bayes factor can be significantly larger.

\vskip 5pt
{\em \textbf{Analysis of GWTC-3 LVK events.}---}
We carried out a Bayesian analysis of selected LVK events to search for scalar field effects.
We analyzed publicly available strain data from the GWTC-3 catalog released by the LVK Collaboration~\cite{LIGOScientific:2018mvr, LIGOScientific:2020ibl, LIGOScientific:2021usb, KAGRA:2021vkt, LIGOScientific:2019lzm, KAGRA:2023pio} using the \texttt{IMRPhenomXPHM\_Scalar} waveform model, performing Bayesian inference using the \textsc{Bilby} package~\cite{Ashton:2018jfp, bilby_pipe_paper}, employing the \textsc{Dynesty} nested sampling algorithm~\cite{Speagle:2019ivv}. In our model, environmental corrections to the waveform are implemented only in the inspiral stage; therefore, analyzing events with low inspiral SNR is expected to be uninformative. Thus, we employ the same event selection criteria that are used in the LVK parameterized tests of general relativity: (i) a false-alarm rate (FAR) smaller than $10^{-3} \,\mathrm{yr}^{-1}$, and (ii) an inspiral SNR greater than 6~\cite{LIGOScientific:2019fpa, LIGOScientific:2020tif, LIGOScientific:2021sio, Roy:2025gzv}. This results in the same event list as in~\cite{Roy:2025gzv}. We do not include binary neutron star events in our analysis, as we have not validated our model for such cases. 
With these criteria, we were left with 28 events to analyze.

Priors for the binary parameters were matched to those used in the LVK analyses. For the additional parameters, we adopted broad uniform priors in $\rhobar$ and $\alpha\leq 0.5 (1+q)$, and constrained the mass of the scalar structure to be less than $10\%$ of the binary's total mass, i.e., $\rhobar r_\phi^3<0.1 M$.
The 90\% upper bounds on the scalar field density for all GWTC-3 events are reported in the SM (Fig. \ref{fig:gwtc_full}), with several lying many orders of magnitude below typical superradiant-cloud densities. For most events, the posteriors are compatible with vacuum, with strong support near zero density. {The exceptions are GW190814 and GW190728, where vacuum lies outside the $90\%$ credible region (see SM).}

\vskip 5pt
{\em \textbf{Superradiance interpretation.}---}
%
A meaningful statistical comparison between the vacuum and scalar-field hypotheses requires physically motivated priors, since the Bayes factor is prior-dependent. It is also essential to exclude unphysical regions of parameter space, where good fits could arise from implausible scalar field densities or particle masses.
We therefore reanalyze GW190814 and GW190728 under the assumption that the scalars originate from superradiance---a well-motivated scenario that (i) imposes stringent constraints on the parameter space and (ii) allows densities high enough to affect current observations (as seen from the 90\% upper bounds in SM).

If the scalars originate from a superradiant instability triggered by a rapidly spinning BH in the binary, the particle mass must be high enough for efficient superradiance but low enough to avoid annihilation into GWs before merger. Accordingly, we restrict the sampling of parameter space to regions where the superradiant and annihilation timescales, $\tau_{\rm SR}$ and $\tau_{\rm ann}$, satisfy $\tau_{\rm SR} \leq \tau_{\rm d}\leq \tau_{\rm ann}$, with $\tau_{\rm d}$ the delay between BH formation (or last spin-up episode) and merger. This yields a much tighter constraint on $\alpha/(1+q)$ than in the previous section. We assume the instability is triggered by the primary BH---whose cloud is more likely to survive until the late inspiral---into the dominant $\ket{211}$ state, which grows faster and to higher densities. Explicit expressions for $\tau_{\rm SR}$ and $\tau_{\rm ann}$ are given in the SM. We consider delay times $\tau_{\rm d}$ between $10^5$ and $10^9$ years, encompassing the timescales of AGN~\cite{Bartos:2016dgn, Mckernan:2017ssq, Yang:2019cbr, Tagawa:2019osr}, dense star clusters~\cite{DiCarlo:2019pmf}, and isolated binary~\cite{Marchant:2016wow, vanSon:2021zpk,Fishbach:2021mhp} formation channels, as previously used in LVK constraints on superradiant clouds (e.g.,~\cite{Aswathi:2025nxa}).

Superradiance further constrains the cloud mass to satisfy $M_\phi/m_1\lesssim \alpha/(1+q)$~\cite{Brito:2015oca}, saturated for a maximally spinning BH and a noninteracting scalar. This sets a theoretical upper bound on the late-inspiral average density \rhobar, realized if the previous saturation mass is fully transferred into the $\ket{100}$ molecule state. 
Tighter bounds would require a detailed understanding of the binary's evolutionary history, the survival fraction of the cloud to the late inspiral, and the mass transferred to monopole states via Bohr resonances (e.g.,~\cite{DellaMonica:2025zby, Tomaselli:2025jfo}).

Applying the above constraints to the priors of the previous section yields prior distributions that (i) are much narrower on $\mb$, increasingly so for larger $\tau_{\rm d}$, and (ii) strongly favor zero scalar-field density. 
Our reanalysis shows that GW190814 provides only weak evidence for a scalar environment ($\lnB \lesssim 1$), whereas GW190728 favors it with $\lnB \approx (3.4,\, 3.5,\, 3.5,\, 2.8)$ for $\tau_{\rm d}[\yrs] = (10^5,\, 10^6,\, 10^7,\, 10^8)$.

\begin{figure}[t]
\includegraphics[width=\linewidth]{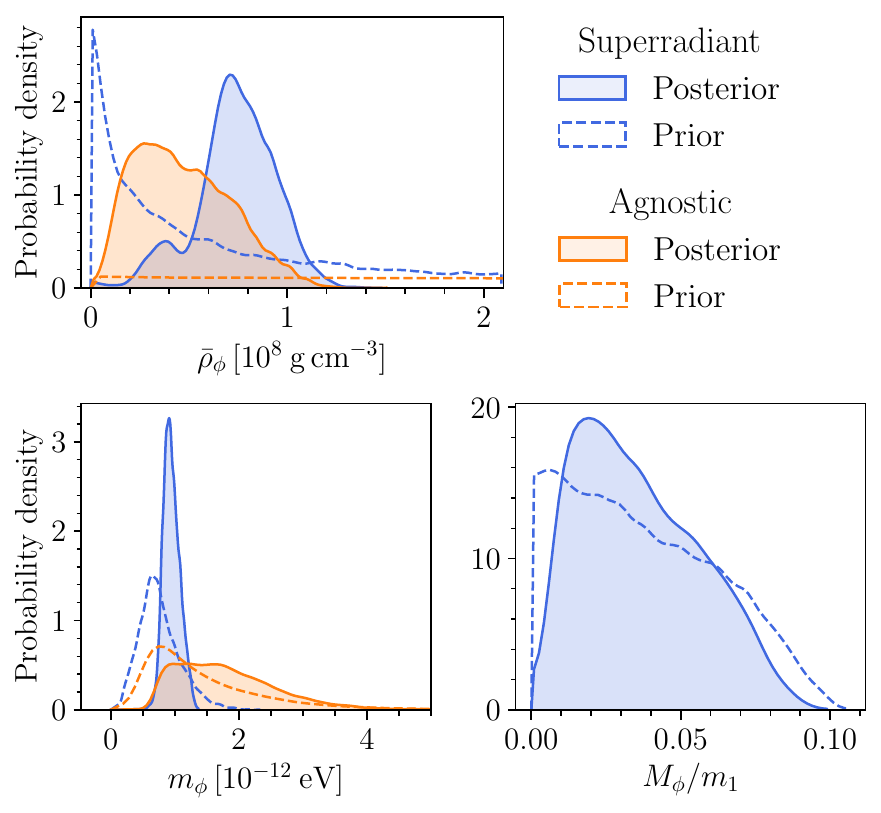}
\caption{Posteriors on scalar field density \rhobar, particle mass~$\mb$, and scalar-to-BH mass-ratio $M_\phi/m_1$ from the analysis of event GW190728, with $\tau_{\rm d}=10^6\,\rm yrs$. {The blue (orange) curves correspond to the superradiance (agnostic) interpretation. The dashed lines show the prior distributions.} 
}
\label{fig:golden_plot}
\end{figure}

\Cref{fig:golden_plot} shows the posteriors on the scalar-field parameters for $\tau_{\rm d}=10^6\,\rm yr$ (other $\tau_{\rm d}$ are shown in SM); $M_\phi$ denotes the scalar-field mass in the late inspiral assuming all of it is in the $\ket{100}$ state. Despite priors favoring vacuum, posteriors peak at nonzero density, and the data prefer the scalar-field model. Superradiance is very restrictive on $\mb$, but our posteriors are still informative: this interpretation points to the existence of a new light scalar with $\mb \sim 10^{-12}\,\mathrm{eV}$.  
The corresponding posteriors on chirp mass and effective spin deviate significantly from those of the vacuum model (\cref{fig:pe:GW190728Contour} in SM).

\vskip 5pt
%
{\em \textbf{Discussion.}---}
%
The existence of new light scalars in strong-gravity fields has so far been constrained by spin measurements of BHs~\cite{Arvanitaki:2014wva, Cardoso:2018tly, Hoof:2024quk, Witte:2024drg, Ng:2020ruv, Aswathi:2025nxa, Caputo:2025oap}, by the nondetection of quasimonochromatic GWs~\cite{Brito:2017zvb, Isi:2018pzk, Palomba:2019vxe, Yuan:2022bem, Collaviti:2024mvh}, and through orbital effects on wide binaries~\cite{GRAVITY:2021xju, GRAVITY:2023cjt, Tomaselli:2025zdo, Branco:2025znp}. Here, for the first time, we probe scalar-field environments through the phasing of GW signals from compact binaries. 
To this end, we developed a waveform model incorporating scalar-field effects, validated it against NR simulations, and applied it in Bayesian analyses of selected GWTC-3 events.
We obtain the first bounds on scalar-field environments of compact binaries. With superradiance priors, GW190728 shows tentative evidence ($\lnB \approx 3.5$), consistent with a light scalar of mass $\mb \sim 10^{-12}\,\mathrm{eV}$.

The high spins of BHs in X-ray binaries, inferred from continuum-fitting and reflection methods, have been interpreted as excluding scalars in this mass range~\cite{Arvanitaki:2014wva, Cardoso:2018tly, Hoof:2024quk, Witte:2024drg}. These spin estimates, however, are highly model-dependent~\cite{Zdziarski:2023zuh, Zdziarski:2024zfg} (see also~\cite{Witte:2024drg}); for instance, accounting for a warm Comptonization layer in Cyg X-1 can reduce the inferred spin parameter to $a/M\lesssim 0.1$~\cite{Zdziarski:2024zfg}.\footnote{Such systematics may also reconcile the apparently high spins of BHs in X-ray binaries with the low spins inferred for most GW sources, allowing for a common origin~\cite{Belczynski:2021agb}.} Constraints from few well-measured BH spins through GWs~\cite{Ng:2020ruv, Aswathi:2025nxa, Caputo:2025oap} do not exclude the full range favored by our analysis of GW190728; the main tension arises from GW190517, whose spin measurement disfavors $\mb \lesssim 1.1\times 10^{-12}\,\mathrm{eV}$. Notably, joint spin measurements from the full GWTC-2~\cite{Ng:2020ruv} showed weak evidence for a light scalar in the mass range suggested by our analysis.

We cannot exclude degeneracies with other vacuum or astrophysical environmental effects. Nevertheless, in GW190728 we found no substantial evidence for eccentricity~\cite{Gamboa:2024hli} or line-of-sight acceleration~\cite{Vijaykumar:2023tjg}, and the data remained consistent with vacuum when allowing for parameterized post-Newtonian deviations in the inspiral waveform, as routinely tested by the LVK~\cite{LIGOScientific:2020tif}. 
We note that this event was highlighted in a search for tidal deformability in LVK events~\cite{Chia:2023tle}; still, no evidence for a non-vanishing tidal deformability was found.
Future observing runs and next-generation ground-based detectors, such as the Einstein Telescope~\cite{ET:2019dnz} and Cosmic Explorer~\cite{Reitze:2019iox}, will deliver louder signals enabling stringent tests for ultralight bosons. Our analysis offers a novel complementary approach toward these future searches.

\vskip 5pt
\noindent
\textbf{\textit{Acknowledgements.}}— We acknowledge useful conversations with Richard Brito, Yifan Chen, Charlie Hoy, Sebastian von Hausegger, and Nico Yunes. We especially thank Juan Calderon-Bustillo, Justin Janquart, Elisa Maggio, and Nicolás Sanchis-Gual for giving feedback on the manuscript.
SR is supported by the Fonds de la Recherche Scientifique - FNRS (Belgium).
RV gratefully acknowledges the support of the Dutch Research Council
(NWO) through an Open Competition Domain Science-M grant, project number OCENW.M.21.375.
JCA acknowledges funding from the Department of Physics at MIT through a CTP Postdoctoral Fellowship.
KC acknowledges support from the Simons Foundation International and the Simons Foundation through Simons Foundation grant SFI-MPS-BH-00012593-03, a UKRI Ernest Rutherford Fellowship (grant number ST/V003240/1) and an STFC Research Grant ST/X000931/1 (Astronomy at Queen Mary 2023-2026).
PGF acknowledges support from the Beecroft Trust and STFC.
We thank the GRTL collaboration (\url{www.grtlcollaboration.org}) for their support and code development work. The authors are grateful for computational resources provided by the LIGO Laboratory and supported by National Science Foundation Grants PHY-0757058 and PHY-0823459. The authors are also grateful for the computational resources provided by Cardiff University supported by STFC grant ST/I006285/1.
This work used Stampede3 at the Texas Advanced Computing Center through allocation PHY250054 from the Advanced Cyberinfrastructure Coordination Ecosystem: Services \& Support (ACCESS) program \cite{ACCESS}, which is supported by U.S. National Science Foundation grants 2138259, 2138286, 2138307, 2137603, and 2138296. The authors acknowledge the Texas Advanced Computing Center (TACC) at The University of Texas at Austin for providing computational resources that have contributed to the research results reported within this paper. URL: \url{www.tacc.utexas.edu}.
This work also used the DiRAC Memory Intensive services Cosma8 and Cosma7 at Durham University, managed by the Institute for Computational Cosmology on behalf of the STFC DiRAC HPC Facility (www.dirac.ac.uk). The DiRAC service at Durham was funded by BEIS, UKRI and STFC capital funding, Durham University and STFC operations grants. DiRAC is part of the UKRI Digital Research Infrastructure.
This research has made use of data or software obtained from the Gravitational Wave Open Science Center (gwosc.org), a service of the LIGO Scientific Collaboration, the Virgo Collaboration, and KAGRA. This material is based upon work supported by NSF's LIGO Laboratory which is a major facility fully funded by the National Science Foundation, as well as the Science and Technology Facilities Council (STFC) of the United Kingdom, the Max-Planck-Society (MPS), and the State of Niedersachsen/Germany for support of the construction of Advanced LIGO and construction and operation of the GEO600 detector. Additional support for Advanced LIGO was provided by the Australian Research Council. Virgo is funded, through the European Gravitational Observatory (EGO), by the French Centre National de Recherche Scientifique (CNRS), the Italian Istituto Nazionale di Fisica Nucleare (INFN) and the Dutch Nikhef, with contributions by institutions from Belgium, Germany, Greece, Hungary, Ireland, Japan, Monaco, Poland, Portugal, Spain. KAGRA is supported by Ministry of Education, Culture, Sports, Science and Technology (MEXT), Japan Society for the Promotion of Science (JSPS) in Japan; National Research Foundation (NRF) and Ministry of Science and ICT (MSIT) in Korea; Academia Sinica (AS) and National Science and Technology Council (NSTC) in Taiwan. We have used \numpy~\cite{Harris:2020xlr}, \scipy~\cite{Virtanen:2019joe}, and \arviz~\cite{arviz_2019} for analyses, and \matplotlib~\cite{Hunter:2007ouj} for preparing the plots in this manuscript.

\bibliography{refs.bib}

\clearpage
\newpage

\section*{Supplemental Material}\label{supp}

\subsection*{Analytic approach to the ionization of\\ gravitational molecules}
In this section, we explain in detail our analytic computation of the binary torque from the interaction with the scalar-field environment. 
We will focus on circular orbits, as we will be interested in the last stages of the inspiral, where eccentricity is expected to have been damped by GW radiation-reaction.

In the non-relativistic approximation, the EKG system~\eqref{eq:EKG} reduces to the simpler Schrödinger-Poisson (SP)
\begin{equation}
	\Big[i  \partial_t  +\frac{\nabla^2}{2 \mb}  +  \mb U\Big]\psi=0 \,, \qquad    \nabla^2U=- 4 \pi  \rho_\phi \,,  \label{eq:SP}
\end{equation}
In the weak field region the dynamical gravitational potential of the binary is
\begin{equation*}
    U\approx\sum_{\ell_*, m_*} \frac{4\pi M}{2 \ell_* +1} Y^*_{\ell_*m_*}(\theta_{2*}, \varphi_{2*}) Y_{\ell_*m_*}(\theta,\varphi)\mathcal{F}_{\ell_*}(r,r_{2*}),
\end{equation*}
with
\begin{align*}
		&\mathcal{F}_{\ell_*}\equiv \tfrac{q_2}{r_{2*}}\left[\big(\tfrac{r}{r_{2*}}\big)^{\ell_*} \Theta(r_{2*}-r)+\big(\tfrac{r_{2*}}{r}\big)^{\ell_*+1}\Theta(r-r_{2*})\right]\nonumber \\ 
        &\quad+\tfrac{(-1)^{\ell_*}q_1}{r_{1*}}\left[\big(\tfrac{r}{r_{1*}}\big)^{\ell_*} \Theta(r_{1*}-r)+\big(\tfrac{r_{1*}}  {r}\big)^{\ell_*+1}\Theta(r-r_{1*}) \right],
\end{align*}
where~$q_j \equiv m_j/M$ is the mass-ratio of the $j$th-component, and~$(r_{j*},\theta_{j_*},\varphi_{j_*})$ its position relative to the center of mass; note that $(r_{1*},r_{2*})=R(q_2,q_1) $, with $R$ the binary's separation,~$q_1=(1+q)^{-1}$ and $q_2=q (1+q)^{-1}$.

Late enough in the inspiral, the binary enters the regime $\beta\equiv R/r_\phi\ll 1$ (with $r_\phi\equiv M/\alpha^2$, and $\alpha\equiv \mb M\ll 1$), which implies a hierarchy in the contribution of binary multipoles to the Schrödinger equation, i.e.,\footnote{Note that the non-relativistic approximation requires~$\Omega\ll \mb$, i.e.,~$(M/R)^2\equiv v^2\ll\beta$, which necessarily breaks close enough to the merger.}
\begin{equation*}
	F_{\ell_*}\approx \frac{q_1 q_2}{R}\Big(\frac{R}{r}\Big)^{\ell_*+1}\left[q_1^{\ell_*}+(-1)^{\ell_*}q_2^{\ell_*}\right]\overbrace{\propto}^{r\to r_\phi} \beta^{\ell+1}\,.
\end{equation*}
This motivates a perturbative approach where, at zero order, the field is expressed as a sum of global (bound and scattering) states of the binary monopole, which is then perturbatively scattered by the higher multipoles of the binary~\cite{Ikeda:2020xvt, Tomaselli:2024ojz, Guo:2025pea}.

At leading order in~$\beta$, the scalar field couples only to the binary's monopole. As the monopole is static and spherically symmetric, we can expand the background scalar as
\begin{align}\label{eq:back_scalar}
	\psi^{(0)}&=\Big(\frac{M_\phi}{\mb}\Big)^{\frac{1}{2}} \Bigl[ \sum_{n, \ell, m}c^{(0)}_{n \ell m} R_{n\ell}(r)Y_{\ell m}(\theta^A)e^{-i \omega t} \nonumber\\
        &\quad+\sum_{\ell, m}\int \frac{\dd k}{2\pi} c^{(0)}_{k \ell m}R_{k\ell}(r)Y_{\ell m}(\theta^A)e^{-i \omega t}\Bigr]\,,
\end{align}
where~$c^{(0)}$ are constants, and the integral is performed over $\omega\in \mathbb{R}^+$. In the regime $\beta\ll 1$, most scalars are located at $r>R$, where the gravitational molecule (global) states look similar to the atom states; i.e., the bound states are
\begin{equation*}
	R_{n\ell}\approx \sqrt{\left( \tfrac{2 \alpha \mb}{n} \right)^3 
		\tfrac{(n - \ell - 1)!}{2n(n + \ell)!}}
	\left( \tfrac{2r}{ n r_\phi} \right)^\ell
	e^{ -\frac{r}{n r_\phi} }
	L_{n - \ell - 1}^{2\ell + 1}\left( \tfrac{2 r}{r_\phi} \right)\,,
\end{equation*}
with~$\omega/\mb \approx -\alpha^2/(2 n^2)$, and the unbound regular states
\begin{equation*}
	R_{k\ell} \approx \frac{2}{r}F_\ell\big(-\frac{1}{k r_\phi},k r\big)\,,	
\end{equation*}
with~$\omega\approx k^2/(2 \mb)$, where $F_\ell$ is the Coulomb function of the first kind. Accretion and the time-dependent higher multipoles lead to the evolution of the scalar field, which is described by~$\dot{c}_{n \ell m}(t)$ and~$\dot{c}_{k \ell m}(t)$.

If the background field has considerable support over $\ell=0$ states, in the regime $\beta\ll1$, these will dominate the energy (angular momentum) exchange with the binary through ionization, because of the suppression $R_\ell(R)\propto \beta^\ell$ for $\ell>0$. Interestingly, in the regime $\beta\ll1$, the torque of this process can be analytically approximated by Eq.~\eqref{eq:torq} where
\begin{align}
     \mathcal{I}_\varphi \approx &\sum_{\ell,m>0} \pi^2 \beta^{\frac{1}{4}+\frac{\ell}{2}}  \left[\frac{2(1+q_2)}{m}\right]^{\frac{1}{2}-\ell} \nonumber \\
     &\quad \times\left[1 +(-1)^\ell q_2^{\ell-1} \right]^2\bigg[\frac{Y_{\ell m}(\frac{\pi}{2},0)}{\Gamma\big(\ell+\frac{3}{2} \big)}\bigg]^2\,,
\end{align}
and $\rhobar\equiv \mb |\psi_{00}^{(0)}(r\ll r_\phi)|^2$, with $\psi_{00}^{(0)}\equiv \int \frac{\dd \Omega}{\sqrt{4\pi}} \psi^{(0)}(\theta^A)$. The last expression can be obtained using the method of variation of parameters as in~\cite{Annulli:2020lyc} (or through Fermi's golden rule~\cite{Baumann:2021fkf, Tomaselli:2023ysb}). In building our waveform model, we employ a cut-off $\ell_{\rm max}= 4$; we verified that our results throughout are independent of $\ell_{\rm max}$, for $\ell_{\rm max}>4$.

\subsection{More on the validation of the waveform model}

\begin{figure}[t]
\includegraphics[width=\columnwidth]{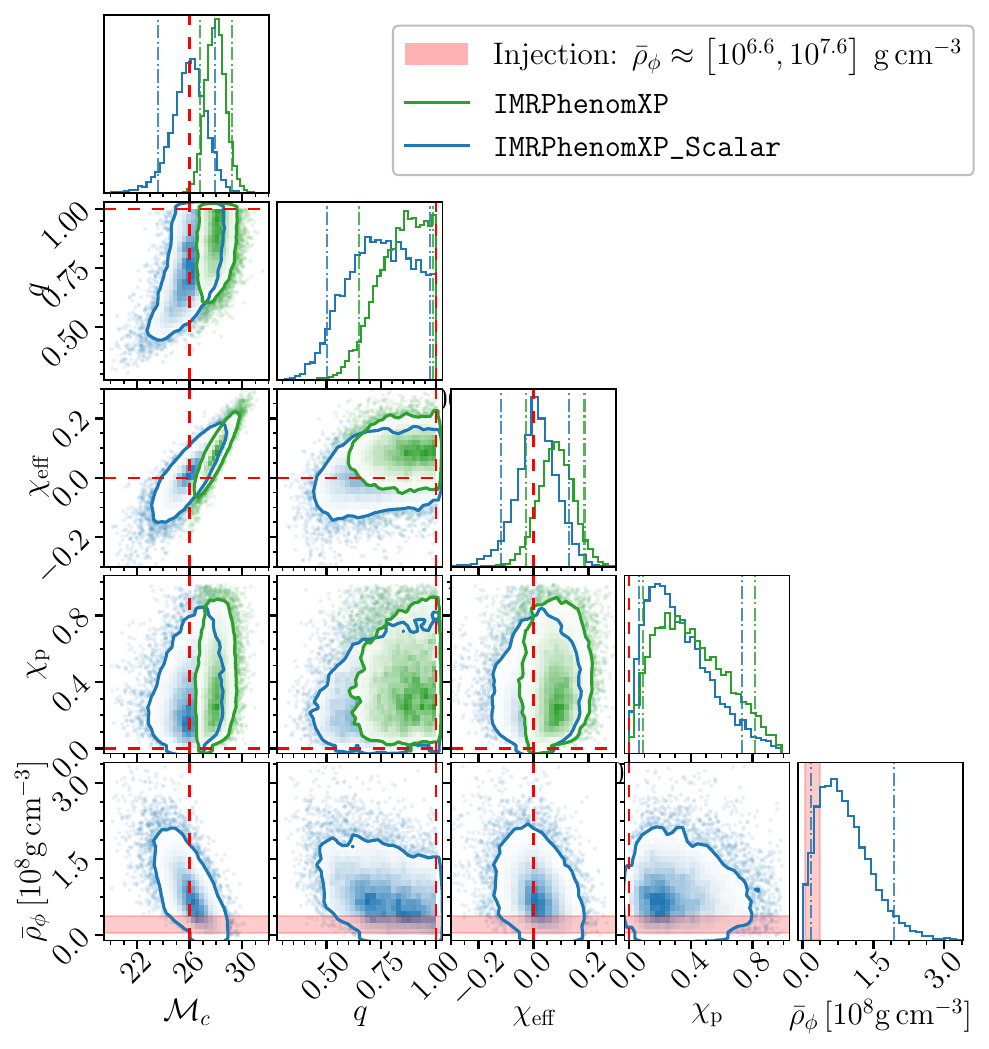}
\caption{Same as~\cref{fig:pe:inj1} (main text) but for a different asymptotic density. The Bayes factor in the model comparison is $\lnB=-0.4$.}
\label{fig:pe:inj2}
\end{figure}

In analyzing the NR waveforms with our Newtonian model, we extrapolate our model considerably beyond its regime of validity, so one should be cautious about its use for several reasons.
First, the NR simulation considered in the main text is not in the regime $\beta\ll1$, where our Newtonian model was derived (the upper panel of \cref{fig:waveform_comparison} shows that the size of the monopolar structure is comparable to the orbital separation). Apart from the breakdown of the multipolar perturbative approach at $\beta \sim 1$, the localization of much of the scalar field in the strong-field region makes the use of Newtonian gravity questionable and the scalar field density is in principle highly gauge-dependent. In \cref{fig:waveform_comparison} we show the scalar field density as measured by the normal observers to the spatial hypersurfaces in the evolution gauge, which do not correspond to more well-defined ones as those in the asymptotically flat region. As explained in the main text, the model is most reliable for systems with~$\alpha\ll1$, where the larger cloud size mitigates the above issues.

We performed Bayesian analyses of the injected NR waveforms to test our waveform model. 
\Cref{fig:pe:inj1} (main text) showed the results of two such analyses for the injected NR waveform corresponding to the simulation of \cref{fig:waveform_comparison} (main text), using the recovery models \phXP and \texttt{IMRPhenomXP\_Scalar}. 

\begin{figure*}[t]
\includegraphics[width=\linewidth]{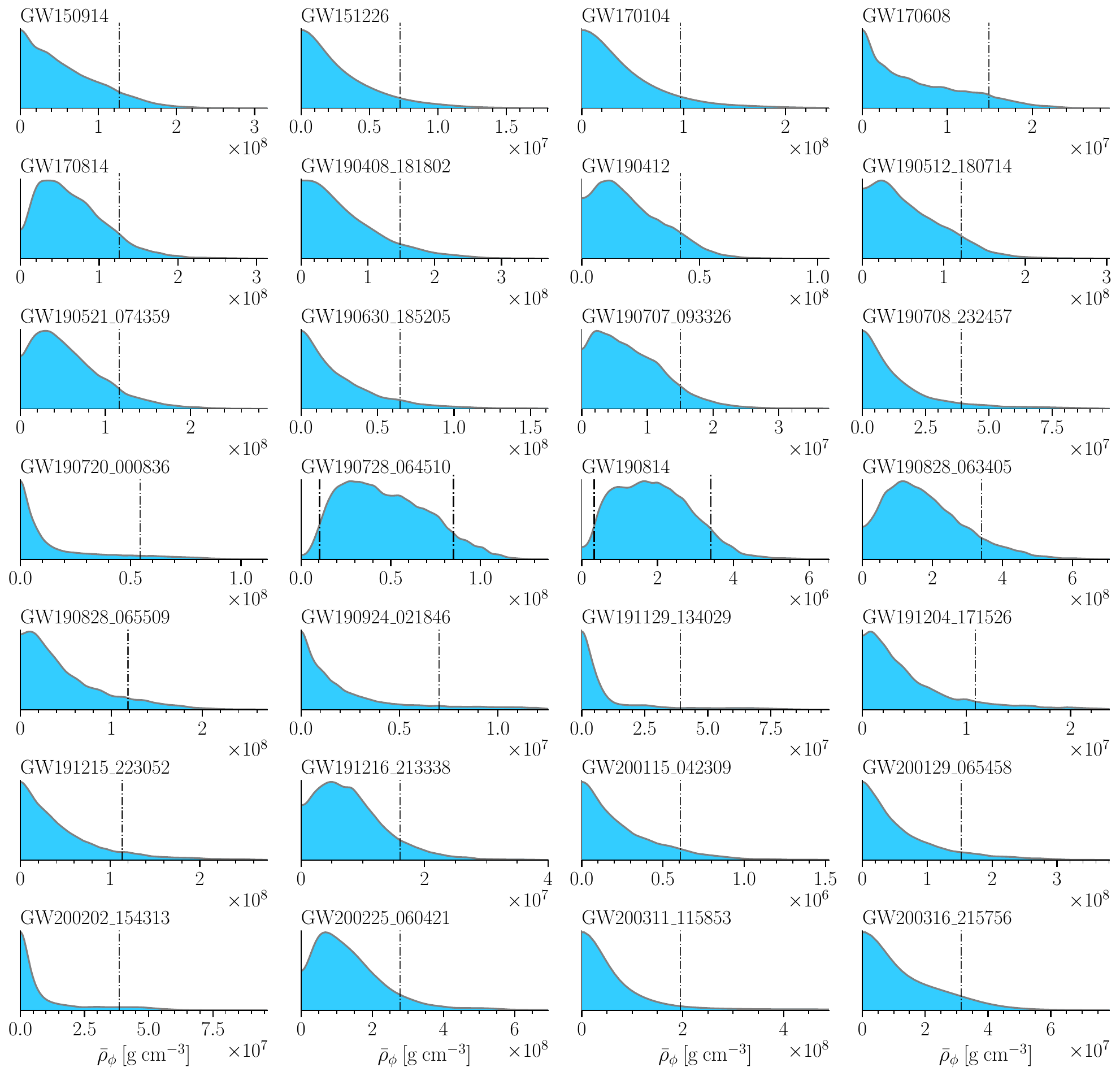}
\caption{Half-violin plots of the scalar-field density posteriors for selected GWTC-3 events. The black dashed line marks the 90\% upper bound ($\rhobar^{90\%}$). Several bounds lie many orders of magnitude below typical superradiant cloud densities.  GW190728 and GW190814 have low support close to zero density, and so we show the 90\% credible interval using the highest density interval method.}
\label{fig:gwtc_full}
\end{figure*}

\Cref{fig:pe:inj2} shows another example. Here, we inject the waveform of an NR simulation with the same parameters as the one in \cref{fig:waveform_comparison} (main text), but with a five-times smaller asymptotic density. In this case, our model still resolves the systematic biases of the vacuum recovery and can accurately estimate the local scalar field density, but finds no evidence for a scalar environment ($\lnB=-0.4$). This likely reflects the Occam penalty from the additional parameter $\alpha$, whose posterior remains uninformative. One could avoid this penalty by performing an analysis fixing the value of $\alpha$, though the results would then be conditional on that choice. We also emphasize that the Bayes factor is sensitive to the adopted prior ranges on density.

\subsection{More on the analysis of GWTC-3 LVK events}
We constrain the effect of light scalars on BH binary dynamics by computing the marginalized posterior probability distribution of the average scalar field density $\rhobar$, integrating the posterior \mbox{$p(\vec{\theta} \mid d,\mathcal{H})$} over nuisance parameters,
\begin{equation}
\label{eq:rho_posterior}
    p\left(\rhobar \,|\, d, \mathcal{H} \right) = \int \left( \prod_{\vec{\theta} \,\setminus \{\rhobar\}} d\theta_i \right) \: p( \vec{\theta} \,|\, d,  \mathcal{H} ),
\end{equation}
where $d$ denotes the data, $\mathcal{H}$ is the hypothesis for a scalar environment, and $\vec{\theta}$ denotes the list of parameters, consisting of the standard binary parameters in vacuum, $\vec{\theta}_{\text{vac}}$, together with two additional parameters characterizing the scalar environment, $\rhobar$ and $\alpha$, such that $\vec{\theta} \equiv \{ \vec{\theta}_{\text{vac}}, \rhobar,  \alpha \}$.

To obtain the posterior distribution \mbox{$p(\vec{\theta} \mid d,\mathcal{H})$}, we typically rely on Markov Chain Monte Carlo or nested sampling algorithms. In this work, we compute the posterior using the \bilby package~\cite{Ashton:2018jfp, bilby_pipe_paper} with the \dynesty nested sampling algorithm~\cite{Speagle:2019ivv}. We use 1000 live points, the sampling method \texttt{acceptance-walk}, an acceptance target of $\texttt{naccept}=60$, and set the log-evidence tolerance to $\texttt{dlogz}=0.1$.

To compare the two competing hypotheses, environment ($\mathcal{H}_{\rm env}$) versus vacuum ($\mathcal{H}_{\rm vac}$), we calculate the ratio of their evidences, known as the Bayes factor, defined as
\begin{equation}
\mathcal{B}_{\rm vac}^{\rm env } = \frac{p\left(d \,|\, \mathcal{H}_{\rm env} \right) }{ p\left(d \,|\, \mathcal{H}_{\rm vac} \right)}
\end{equation}
where $p\left(d \,|\, \mathcal{H}_{\rm env} \right)$ and  $p\left(d \,|\, \mathcal{H}_{\rm vac} \right)$ are the evidences for the environment and vacuum hypotheses, respectively. In this work, we report the natural logarithm of the Bayes factors.

Using the setup described above, we re-analyzed all LVK events of the GWTC-3 catalog satisfying the selection criteria introduced in the main text using our waveform model \texttt{IMRPhenomXPHM\_Scalar}. \Cref{fig:gwtc_full} displays the histograms of the marginalized posterior distribution of the scalar field density $\rhobar$. Most posteriors show significant support near the lower bound of the density prior, consistent with the vacuum hypothesis. 
Two exceptions are GW190728 and GW190814, which show relatively high support at larger density values and yield $\lnB\approx0.4$ and $0.6$, respectively---positive but not significant.

\Cref{fig:gwtc_full} also shows the 90\% upper bounds on the scalar-field density. The tightest limit is from GW200115{\niceunderscore}042309 with $\rhobar \lesssim 10^{5.8}\,\densityUnit$.
Among the 28 analyzed events, four yield 90\% bounds below $10^{7}\, \densityUnit$. The weakest limits are obtained from GW190828{\niceunderscore}063405 and GW200225{\niceunderscore}060421, with $\rhobar \lesssim 10^{8.5} \densityUnit$ and $\rhobar \lesssim 10^{8.4} \densityUnit$, respectively. The weaker bound from GW190828{\niceunderscore}063405 is attributable to its relatively high redshift ($z \sim 0.38$)~\cite{LIGOScientific:2021usb}, which results in fewer inspiral cycles within the sensitive frequency band of the LVK detectors.
{As discussed in the main text, we restricted to~$\alpha\leq 0.5(1+q)$ where our perturbative approach is more reliable; since the coupling parameter $\alpha$ is anti-correlated with the average density~$\rhobar$, removing this restriction would generally lead to tighter (i.e.\ smaller) 90\% upper bounds. Therefore, our treatment yields conservative upper bounds.}

In \Cref{fig:gwtc_full}, we show the 90\% credible intervals computed using the highest-density interval (HDI) method implemented in \arviz~\cite{arviz_2019}. For most events, we notice the lower bound of the 90\% credible interval is consistent with vacuum. However, for GW190728 and GW190814 the density posteriors are barely consistent with vacuum. 

Since a low–density environment is effectively unobservable with LVK sensitivities and the prior range on $\rhobar$ is extremely broad, posteriors may appear offset from zero even when the binary is in vacuum. 
So, one can define a reference density $\rhobar^{\ast}$ (whose value is event dependent), below which environmental effects are indistinguishable from vacuum. We define such reference density as follows.
For each event, we take the maximum-likelihood parameters $\vec{\theta}_{\rm vac}$ from the vacuum analyses reported in the GWTC catalog~\cite{LIGOScientific:2021usb, KAGRA:2021vkt}, and assume a scalar environment with $\alpha$ set to either the lower or upper bounds of our prior (i.e., 0.008 and 0.5). We then compute the mismatch between the vacuum waveform $h(\vec{\theta}_{\rm vac})$ and the environmental waveform $h(\vec{\theta}_{\rm vac},\alpha,\rhobar)$ as a function of $\rhobar$, weighting the inner product using the Livingston PSD. We define $\rhobar^{\ast}$ as the density leading to a mismatch of 1\%.

For GW190814, $\rhobar^\ast$ lies outside the 97.5\% and 90\% credible regions for $\alpha = 0.5$ and $\alpha = 0.008$, respectively. For GW190728, $\rhobar^\ast$ lies outside the 99.7\% credible region for both $\alpha$ values. Notably, for the latter, vacuum lies outside the 99.5\% credible region even when using a reference density corresponding to a 3\% mismatch (such mismatch is expected to generate an observable deviation in LVK data).

\subsection*{More on the superradiance interpretation}

Here, we provide details on the reanalysis of GW190814 and GW190728 under the assumption that the scalars originate from superradiance, discuss the possibility that a significant fraction of the scalar field is found in monopole states during the late inspiral, and compare the resulting binary parameter estimates with those obtained with the vacuum model.

We begin by presenting the explicit constraints on the parameter space imposed in our reanalysis by superradiance (which we refer to as superradiance-motivated priors). The condition $\tau_{\rm SR} \leq \tau_{\rm d} \leq \tau_{\rm ann}$ translates into~\cite{Brito:2015oca}
\begin{equation}\small
    0.02 \Bigl[\frac{m_1}{10 M_\odot} \frac{1\, \rm Myr}{\tau_{\rm d}} \Bigr]^{\frac{1}{9}}\lesssim\frac{\alpha}{1+q}\lesssim 0.11 \Bigl[\frac{m_1}{10 M_\odot} \frac{1\, \rm Myr}{\tau_{\rm d}} \Bigr]^{\frac{1}{15}}\,.
    \label{eq:constrain:alpha}
\end{equation}
The maximum saturation mass, $M_\phi/m_1 \lesssim \alpha/(1+q)$~\cite{Brito:2015oca}, further implies
\begin{equation}\small
    \bar{\rho}_\phi\lesssim \frac{1.3\times 10^8 \densityUnit}{8(1+q)^{-3}} \Bigl[\frac{\alpha/(1+q)}{0.07} \Bigr]^7 \Bigl[\frac{10 M_\odot}{m_1}\Bigr]^2\,,
    \label{eq:constrain:rho}
\end{equation}
which is realized if the saturation mass is fully transferred into the $\ket{100}$ molecule state (i.e., $|c_{100}^{(0)}|=1$ in \cref{eq:back_scalar}).

Perturbative analyses of scalar-field dynamics, using gravitational-atom states as the expansion basis, indicate that---except in a limited region of parameter space (e.g., nearly retrograde binaries)---the scalar cloud is reabsorbed early in the inspiral through resonant (hyper)fine transitions induced by the companion~\cite{Tomaselli:2024bdd, Tomaselli:2024dbw}. These results hold for nearly equal-mass binaries, as the companion is outside the cloud when the transitions are excited. However, in some formation channels, binaries may originate with separations close enough to evade such resonances, and rapid hardening in astrophysical processes (e.g., a common-envelope phase~\cite{Guo:2024iye}) could break the (hyper)fine floating resonances~\cite{Tomaselli:2024bdd} and allow cloud survival.

\begin{figure}[t!]
\includegraphics[width=\columnwidth]{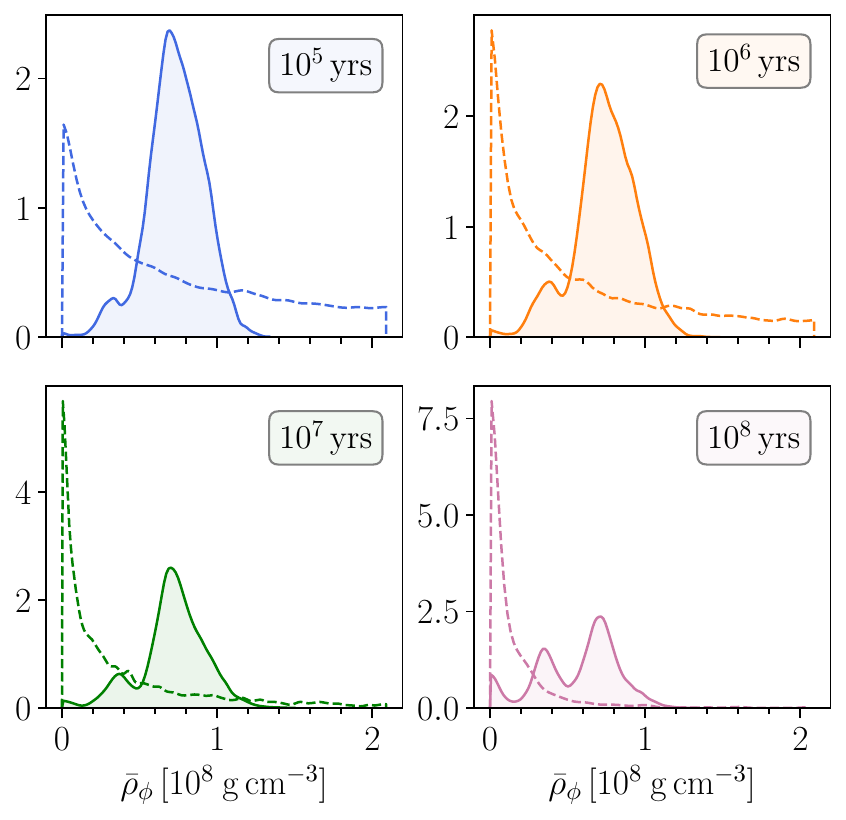}
\caption{Posterior versus prior distributions of the scalar field density \rhobar for GW190728, considering different delay times $\tau_{\rm d}$ between BH formation and merger (cf. superradiant interpretation). Longer delay times favor the vacuum hypothesis more strongly, but the posteriors still support nonzero densities.}
\label{fig:pe:GW190728density}
\end{figure}
\begin{figure}[t]
\includegraphics[width=\columnwidth]{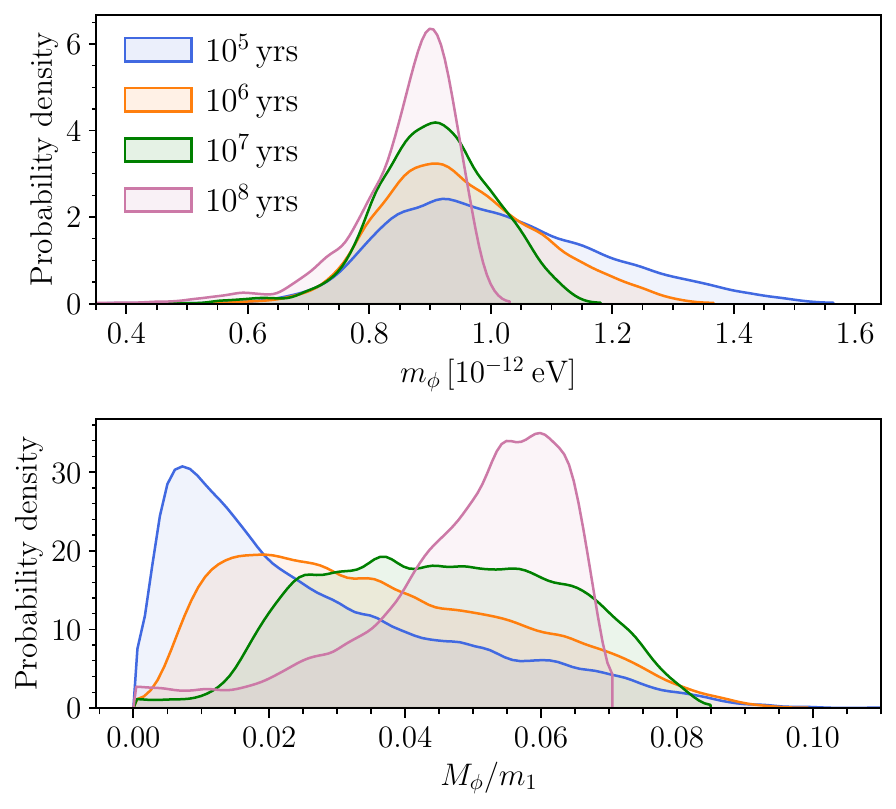}
\caption{Posteriors of the particle mass $m_\phi$ (top panel) and scalar-to-BH mass-ratio $M_\phi/m_1$ (bottom panel) for GW190728 with different delay times $\tau_{\rm d}$ between BH formation and merger (cf. superradiant interpretation).}
\label{fig:pe:GW190728bosons}
\end{figure}

In any case, our results show that even if only a small fraction of the cloud survives to the late inspiral and is transferred into monopole molecule states (e.g., via Bohr floating resonances~\cite{Tomaselli:2025jfo}), the scenario can still be probed by current ground-based GW observations such as those of LVK. For $\tau_{\rm d}\lesssim 10^6\, \yrs$, the analysis of GW190728 is compatible with a late-inspiral cloud mass of a few percent of the maximum saturation value.

Figure~\ref{fig:pe:GW190728density} shows the comparison between the posterior distribution of the scalar field density and its prior for analyses with different timescales $\tau_{\rm d}$. Although the initial prior on the density is uniform, the timescale-based constraints on $\alpha$ and the upper bound on $\rhobar$ (determined using Eqs.~\eqref{eq:constrain:alpha} and \eqref{eq:constrain:rho}) lead to a nonuniform prior, which increasingly supports the vacuum as $\tau_{\rm d}$ grows. Despite that, for a delay time $\tau_{\rm d}\lesssim 10^7 \,\yrs$ the evidence for environmental effects remains $\lnB \approx 3.5$.

\begin{figure}[t!]
\includegraphics[width=\columnwidth]{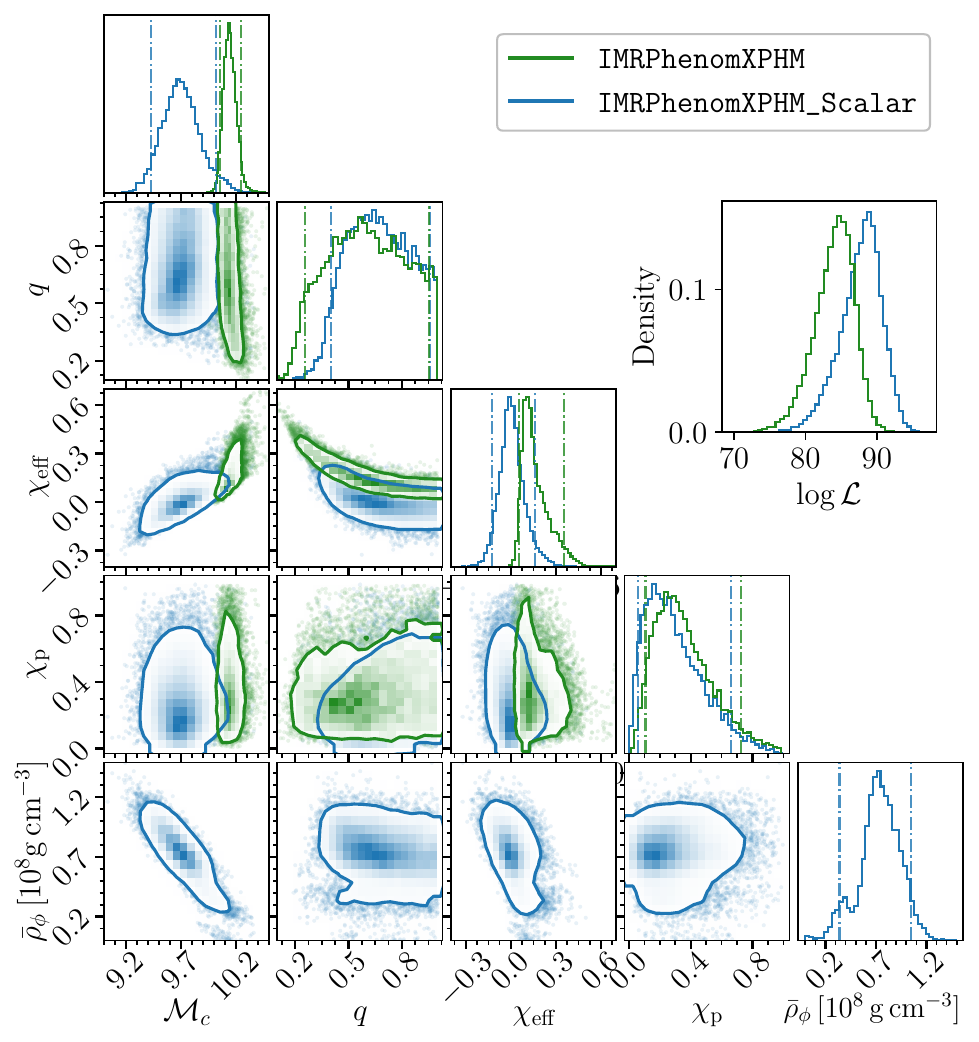}
\caption{Marginalized posterior distributions of GW190728 binary parameters showing deviations between models. The blue plots correspond to the environmental model \texttt{IMRPhenomXPHM\_Scalar}, with $\tau_{\rm d}=10^6 \yrs$, and the green plots correspond to the vacuum model \texttt{IMRPhenomXPHM}. The inset shows the log-likelihood distributions for both analyses.}
\label{fig:pe:GW190728Contour}
\end{figure}

Fig.~\ref{fig:pe:GW190728Contour} shows the marginalized posterior distributions of the GW190728 binary parameters inferred using the environmental model with $\tau_{\rm d} = 10^6 \yrs$, compared against the standard vacuum model. We observe significant deviations in both the chirp mass and the effective spin parameters. Neglecting environmental effects in the analysis leads to a systematically heavier chirp mass and an increased effective spin. The 90\% credible intervals for the chirp mass are $\mathcal{M}_c = 9.69^{+0.27}_{-0.28}M_\odot$ (environmental model) and $\mathcal{M}_c = 10.14^{+0.10}_{-0.09}\:M_\odot$ (vacuum model). Since environmental effects shorten the signal, the chirp mass recovered with the vacuum model is expected to be biased toward larger values. A noticeable bias is also present in the inferred effective spin: the vacuum model strongly supports a nonzero spin, $\chi_{\rm eff} = 0.14^{+0.18}_{-0.11}$, while the environmental model indicates consistency with a nonspinning binary, $\chi_{\rm eff} = 0.0^{+0.15}_{-0.13}$. This bias is consistent with the NR injection analyses shown in Fig.~\ref{fig:pe:inj1} and \ref{fig:pe:inj2}. 

The inset in Fig.~\ref{fig:pe:GW190728Contour} compares the distributions of the log-likelihood values for the vacuum and environmental waveform models.  The distribution of log-likelihood values peaks at higher values for the environmental model, indicating an improved fit to the data. The median log-likelihood difference between the two models is $\Delta \ln \mathcal{L} \approx 3.7$, consistent with  evidence in favor of the environmental model.

To corroborate the evidence for an environmental effect surrounding the GW190728 binary, we further verified the data quality and found no glitch or noise artifact reported by the LVK~\cite{LIGOScientific:2020ibl, LIGOScientific:2021usb}. We also performed separate analyses using data from individual detectors and compared the results against the vacuum model. Evidence for the environmental model is found in both the detectors, LIGO Hanford (LHO) and LIGO Livingston (LLO). The network Bayes factor is largely dominated by the LLO signal, while the evidence from LHO is weaker. This is consistent with the respective signal-to-noise ratios, SNR in LLO $\sim 11$ and SNR in LHO $\sim 7$, and with the relatively lower low-frequency sensitivity of LHO during O3a. Nevertheless, the individual-detector analyses show similar deviation trends in the inferred mass and spins as observed in the network analysis (see Fig.~\ref{fig:pe:GW190728Contour}).

As environmental effects drive binaries to inspiral more rapidly and merge sooner than binaries in vacuum, we expect a difference in the number of cycles within the LVK band starting at a lower cutoff frequency of $20 \Hz$. For the maximum-likelihood sample from the scalar-environment analysis with $\tau_{\rm d} \lesssim 10^6\:\yrs$, the number of in-band cycles is 42, whereas the vacuum waveform has 52 cycles. This difference of 10 cycles could be significant, depending on the detector sensitivity at low frequencies. For the same maximum-likelihood parameters, we find the number of cycles to be 30/35, 22/25, and 17/18 for environment/vacuum when the starting frequency is 25, 30, and $35\Hz$, respectively. This implies that the environmental contribution becomes negligible above $35 \Hz$. To ensure consistency, we further perform analyses varying the lower cutoff frequency for both the environment and vacuum models. We find similar evidence for environmental effects for a starting frequency at $25 \Hz$ and $30\Hz$, with $\lnB \gtrsim 3.5$ and the density posterior strongly supporting non-zero values, but the evidence becomes negligible when the lower cutoff is increased to $35\Hz$.

\end{document}